# PHASE DOMAINS AND SPATIAL SOLITONS IN DEGENERATE OPTICAL PARAMETRIC OSCILLATORS WITH INJECTION

**Kestutis Staliunas**

Physikalisch Technische Bundesanstalt, 38116 Braunschweig, Germany

and

**Víctor J. Sánchez-Morcillo**

Departament D'Òptica, Universitat de València, Dr.Moliner 50, E-46100 Burjassot, Spain

## Abstract

The stability of phase domains and spatial solitons in DOPO under the presence of an injected signal is investigated. The injected signal prevents the nondegenerate regime and, for a particular value of the phase, preserves the equivalence between the two homogeneous states, allowing the domain formation and, in particular, the stability of solitons. The main conclusion is that injection facilitates the experimental observation of solitons in degenerate OPOs.

The radiation in Degenerate Optical Parametric Oscillators (DOPOs), and also in some other systems such as degenerate four-wave mixing, prefers two values for the phase (differing by π). These fixed phase systems are characterized by a pitchfork bifurcation, which has profound consequences in their pattern formation properties: the order parameter is real-valued, and patterns such as stripes, hexagons, or phase domains are favored. In contrast, the order parameter in lasers, nondegenerate optical parametric oscillators (OPOs), *e.a.*, is complex-valued, and patterns as tilted waves, optical vortices and vortex lattices are excited.

The phase domains in DOPOs are the areas in the space characterized by relatively constant intensity, and one of two values of the phase, as predicted in [1]. The domains are separated by domain walls, which are dark lines with zero light intensity. Depending on the DOPO resonator detuning, the domain boundaries can display different dynamics [2,3]. For small and zero detunings, when the DOPO emission is predominantly along the optical axis, the domains shrink, and eventually disappear. For large (negative) detunings, when the DOPO emits off-axis, the domains grow, and eventually transform into labyrinths, or stripe patterns (with defects). For intermediate detunings the shrinking domains can stabilize at particular radii, constituting dark ring type spatial localized structures or solitons.

The domain dynamics and especially the solitons were recently intensively investigated in DOPOS [4], and in some other nonlinear optical systems displaying pitchfork bifurcation [5].

Experimentally the phase domains and the ring type solitons were demonstrated in degenerate four-wave mixing [6], but not yet in DOPOs.

The experimental realization of the theoretically predicted phenomena in DOPOs discussed above possess some problems. One of such problems is the realization of degenerate regime, and of single-longitudinal-mode emission. The realization of frequency-degenerate regime emission is necessary, since the phase domains and ring type solitons rely on the real-valued nature of the order parameter and the associated pitchfork bifurcation. In OPOs conversely the order parameter is complex-valued, and the bifurcation is of Hopf type [7].

OPOs usually have a very broad spectral width of phase synchronization. This means, that whereas the energy conservation imposes a condition on the sum of the frequencies $w_1 + w_2 = w_3$, where $w_3$ is the frequency of the pump wave, the individual generated frequencies $w_1$ and $w_2$ have a freedom to vary in a large parameter range. If the longitudinal mode separation of OPO resonator is $dw$, then not only the degenerate regime $w_1 = w_2 = w_3/2$ is allowed, but also nondegenerate emission, with frequencies $w_1 = w_3/2 + n\, dw$, $w_2 = w_3/2 - n\, dw$, being $n$ an integer. In experiments this can cause a hopping among the longitudinal modes (with different *n*), a hopping between degenerate and nondegenerate emission or, a multimode emission.

The easiest solution of this problem would be the seeding of the OPO, by injecting a signal exactly of the half of the pump frequency. However, the injection also modifies the stability of the spatial patterns, and it is not clear whether this seeding will not destroy the expected solitons. It is also well known that even a very weak injection into lasers breaks the phase invariance and can lock the phase of the laser to the phase of injection laser. One can expect that the injection will break the equivalency between the two branches of the pitchfork bifurcation of the DOPO. However, the symmetry between two equivalent solutions, differing by π in phase, is crucial for the phase domain formation, and for the stability of dark ring type spatial solitons.

In this letter we study the effects of the injection in a DOPO. We show, that the injection with the correct phase (*p/2* or -*p/2*) indeed does not break the equivalence between the two branches of the pitchfork bifurcation. It thus allows the formation of phase domains, and spatial solitons of dark-ring shape, such as predicted in [4]. On the other hand, it simplifies strongly the problem of experimental observation of the predicted patterns, since the injection fixes the frequency of the radiation and prevents the nondegenerate emission.

We consider a doubly resonant DOPO with signal injection, where both the subharmonics $A_1(r,t)$ and the pump wave $A_0(r,t)$ are close to a cavity resonance [8]

$$\partial_t A_1 = g_1[Y - (1 + i\Delta_1)A_1 + A_1^* A_0 + i a_1 \nabla^2 A_1], \qquad (1a)$$

$$\partial_t A_0 = g_0[E - (1 + i\Delta_0)A_0 - A_1^2 + i a_0 \nabla^2 A_0], \qquad (1b)$$

where $Y = y\, exp(ij)$ is the complex amplitude of the injected signal, *E* is the amplitude of the (external) pumping field, $D_1$ and $D_0$ are the detunings of the resonators, $g_1$ and $g_0$ are the decay parameters, and $a_1$ and $a_0$ are the diffraction coefficients.

Due to the complexity of the solutions introduced by the injection, we perform a numerical study. However, some insight on the role of injection can be gained by analyzing a simplified model for DOPOs, the parametrically driven Ginzburg-Landau equation (PGLE)

$$\partial_t A_1 = Y + E A_1^* - (1 + i\Delta_1)A_1 + A_1|A_1|^2 + ia_1 \nabla^2 A_1. \qquad (2)$$

The PGLE without signal injection has been derived for DOPOs in [9,10]. The constant term in (2) can be considered to be added phenomenologically. Also, Eq.(2) can be easily obtained by repeating the adiabatic elimination [9] or multiscale expansion [10] derivation procedures used to obtain (2) without the signal injection.

Eq.(2) is variational ($\partial_t A_1 = -dF/dA_1^*$) for zero signal detuning $\Delta_1 = 0$, and neglecting the spatial dependence of the fields ($a_1 = 0$). The analysis of its variational potential allows understanding the role of the injection. The potential, whose minima correspond to the stable states of the system, is

$$F = -y\left(A_1 e^{-i\varphi} + A_1^* e^{i\varphi}\right) - E\frac{A_1^2 + A_1^{*2}}{2} + |A_1|^2 + \frac{1}{2}|A_1|^4, \qquad (3)$$

and is plotted in Fig.1 for different values of the injection.

Without the injected signal, $y = 0$, the potential has two minima [Fig.1(a)], corresponding to two stable solutions of DOPO (1) as well as the PGLE (2). These minima are placed symmetrically with respect to the zero solution, which corresponds to a saddle point. The phase domains correspond to realizations of solutions corresponding to potential minima, and the domain boundaries to homoclinic connections between these two stable points.

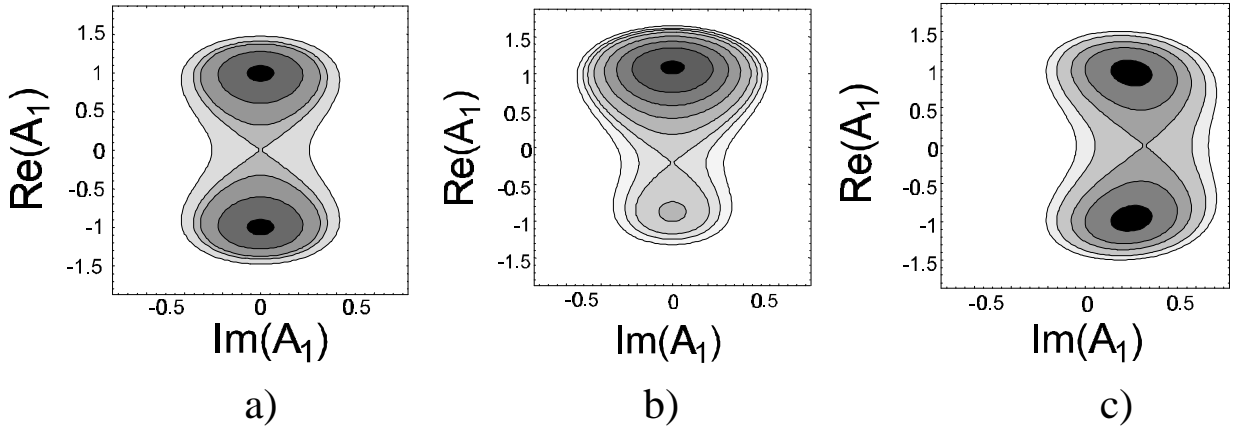

**Fig.1.** *Isopotential curves depending on the injection. (a) Y = 0, (b) Y = 0.1 ($\varphi = 0$), and (c) Y = i ($\varphi = \pi/2$). The pump value is E = 2. The potential minima correspond to black regions.*

Injection modifies the potential in such a way that, in general, it breaks the symmetry of the phase space, as Fig.1(*b*) shows. Although, for small values of the injection both two states may coexist, however the attraction basin of one solution is larger (the potential is deeper) than for the other. Therefore, when starting from a weak (quantum) noise, the eventual result is the realization of the homogeneous solution corresponding to the state having a deeper potential minimum

The equivalence (symmetry) between states, required for domain formation, is however preserved in the special case $\varphi = \pm \pi/2$. Fig.1(*c*) shows this case. The potential minima, as well as the saddle point, are shifted horizontally due to the injection. As the result, the two phases remain equivalent. Also, the zero solution is on the boundary between the attraction basins of both stationary points. Therefore the DOPO with injection, started from quantum noise, can lead to domain structures.

With these preliminary conclusions derived from the potential picture of DOPOs, the spatio-temporal dynamics of the domains, and also the stability of dark ring type localized structures were investigated by integrating numerically the full DOPO equations (1). The result is the behavior regions in the parameter space $\langle \Delta_1, y \rangle$ plotted in Fig.2. For $y = 0$, the diagram reproduces results given in [2]: for positive signal detunings the dark ring collapses, while for too large negative detunings the spatial soliton nucleates new solitons, and a (seemingly transient) hexagonal pattern occurs. For small injection, the stability domain of solitons is weakly affected: positive (negative) injections $y$ shift the soliton stability range to the negative (positive) values of signal detuning. This asymmetry of the soliton existence range with respect to the sign of injection $y$ is due to the fact, that the phases of spatial domains are in general not precisely equal to 0 and $\pi$, but depend on the signal detuning. In particular, for $y = 0$:

$$\varphi_{-} = \frac{1}{2}arcsin(-\frac{\Delta_1}{E}), \; \varphi_{+} = \varphi_{-} + \pi. \tag{4}$$

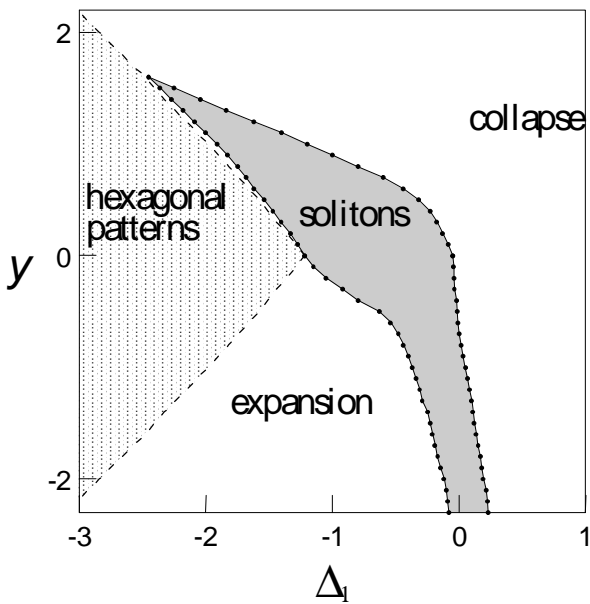

*Fig2*. Stability domain of ring solitons, in the plane $\langle \Delta_1, y \rangle$, evaluated from numerical integration if (1). The other parameters are $E = 2$, $a_1 = 2a_0 = 0.0005$, $g_0 = g_1 = 1$, $\Delta_0 = 0$ and $\varphi = \pi/2$. The phase of the background field is $\varphi_{back} \approx 0$ (4); the phase at the center of the soliton is opposite, $\varphi_{sol} \approx \pi$.

The domains then become not precisely equivalent at the injection with the phase $\varphi = \pm\pi/2$. E.g., for positive sign of injection $y > 0$ supports the background solution (4) for $y$ small, and attenuates the field inside the dark ring. For negative sign of injection $y < 0$ the result is opposite. Soliton dynamics is more complicated for larger values of the injection. Depending on the injection, the solitons may loose the stability in different scenarios for negative detuning values (left boundary of the soliton stability diagram in Fig.2). We observe either the nucleation of new solitons for $y > 0$, leaving an extended pattern with hexagonal symmetry (this transition corresponds to the Turing instability of the homogeneous solution), or expansion of solitons to large domains for $y < 0$, leaving again an homogeneous solution but with the opposite phase, the one corresponding to the inner region of the ring. In contrast, for positive detunings (right boundary of the soliton stability diagram in Fig.2) the solitons always loose their stability through a collapse of the ring (leaving a homogeneous solution with the dominant phase, corresponding to that outside of the ring), for any sign of the injection. As seen from Fig.2, an injection with small amplitude (few percent of the amplitude of the signal radiation) affects the existence range of the solitons very weakly. The solitons do exist even for amplitudes of the injection almost as large as the signal field in the cavity, however the soliton stability range is strongly modified with respect to injectionless case.

In conclusion, we have shown that an injected signal in an OPO, on one hand, favors the experimental realization of the degenerate OPO emission. On the other hand, an appropriate selection of the phase of the injection ($\varphi = \pm\pi/2$), maintains the existence of spatial solitons, opening a way for their experimental observation.

We acknowledge discussions with C.O. Weiss, V. Sirutkaitis, G. Slekys and G. J. De Valcárcel. This work has been supported by Acciones Integradas project HA1997-0130, NATO grant HTECH.LG 970522, Sonderforschungs Bereich 407 and by DGICYT of the Spanish Government under project n PB98-0935-C03-02

**References:**

1.  S. Trillo, M. Haelterman and A. Sheppard, Opt. Lett. **22**, 970 (1997).
2.  K. Staliunas and V.J. Sánchez-Morcillo, Phys. Rev. A **57**, 1454 (1998).
3.  K. Staliunas and V.J. Sánchez-Morcillo, Phys. Lett. A **241**, 28 (1998).
4.  G.L. Oppo, A.J. Scroggie and W.J. Firth, J. Opt. B: Quantum Semiclass. Opt. **1**, 133 (1999); M. Le Berre, D. Leduc, E. Ressayre and A. Tallet, J. Opt. B: Quantum Semiclass. Opt. **1**, 153 (1999).
5.  M. Tlidi, P. Mandel, R. Lefever, Phys. Rev. Lett. **73**, 640 (1994).
6.  V.B. Taranenko, K. Staliunas and C.O. Weiss, Phys. Rev. Lett. **81**, 2236 (1998).
7.  G.L. Oppo, M. Brambilla, D. Camesasca, A. Gatti and L.A. Lugiato, J. Mod. Opt **41**, 1151 (1994).
8.  G.L. Oppo, M. Brambilla and L.A. Lugiato, Phys. Rev. A **49**, 2028 (1994).
9.  K. Staliunas, J. Mod. Opt **42**, 1261 (1995).
10. S. Longhi, Phys. Scr. **56**, 611 (1997).